\documentclass[11pt]{article}
\usepackage[dvips]{epsfig}
\usepackage[T1]{fontenc}
\usepackage[latin1]{inputenc}
\usepackage{graphicx}
\usepackage[english]{babel}
\usepackage{amsmath}
\usepackage{amssymb}
\usepackage{amsfonts}
\usepackage[T1]{fontenc}
\usepackage{color}
\usepackage{babel}
\usepackage{verbatim}
\usepackage[unicode=true,pdfusetitle,bookmarks=true,bookmarksnumbered=false,bookmarksopen=false,
 breaklinks=false,pdfborder={0 0 1},backref=false,colorlinks=true]{hyperref}
\hypersetup{linkcolor=blue,citecolor=blue}
\makeatletter
\usepackage[dvips]{epsfig}
\usepackage[T1]{fontenc}
\usepackage{color}
\usepackage{pdflscape}
\usepackage{cite}

\begin{document}

\title{\bf Non-Vacuum Solutions in Cotton Theory}
\author{Serkan Doruk Hazinedar$^a$\thanks{Email: doruk.hazinedar@bilkent.edu.tr},~
    Yaghoub Heydarzade$^a$\thanks{Email: yheydarzade@bilkent.edu.tr}, and Maryam Ranjbar$^b$\thanks{Email: maryamrnjbr96@gmail.com} \\
    \\
    {\small $^a$Department of Mathematics, Faculty of Sciences, Bilkent University, 06800 Ankara, Turkey} \\ {\small $^b$Department of Physics and Astronomy, University of California, Riverside, California 92521, USA}}
\date{\today}
\maketitle
\begin{abstract}
Cotton Theory (CT) introduces a higher derivative extension of General Relativity (GR) characterized by third-rank field equations. Recently, key distinctions between CT and GR concerning wave and vacuum solutions have been highlighted in \cite{Gurses, Altas}. In this study, two particular non-vacuum solutions of CT are investigated within its Codazzi formulation. The motivation is to reveal how this theory might account for or adapt to the effects of non-vacuum sources, and whether it can provide new insights into the behavior of both singular and regular black holes in astrophysical contexts. It is shown that CT generalizes the Kiselev and Dymnikova solutions in GR. Some aspects of the generalized solutions, in particular concerning singularities, thermodynamics, and geodesics, are addressed in comparison to GR.
        \end{abstract}


\section{Introduction}
The Cotton tensor, named after Emile Cotton \cite{Cotton}, and Weyl tensor, named after Herman Weyl \cite{Weyl}, are 3rd-rank and 4th-rank curvature tensors that describe the curvature of Riemannian manifold of dimension $n$, respectively. The Weyl-Schouten theorem \cite{Eisenhart} states that for any $n$-dimensional pseudo-Riemannian manifold
when $n \geq 4$, the manifold is conformally flat if and only if its Weyl tensor vanishes, and
when $n = 3$, the manifold is conformally flat if and only if its Cotton tensor vanishes. Hence, these tensors have a significant role in conformal structures of manifolds in  $3$-dimensional and higher dimensional gravity theories, see for instances \cite{deser, manheim1, manheim2}. 
Recently, a new geometric theory of gravity based on the Cotton tensor was introduced by Harada \cite{Harada}. This theory is called the Cotton Theory of Gravity (CT) and possesses third-rank tensorial field equations. The field equations of CT can be derived from the conformal Weyl action \cite{manheim1}  varying the connection while keeping the metric fixed. A vacuum solution of CT for a static spherically symmetric spacetime is a Schwarzschild-like solution \cite{Harada}. Utilizing this solution,
 it was shown in \cite{Harada2} that CT is capable of addressing the galaxies' rotation curve problem in GR. Later,  general classes of spherically symmetric solutions of CT were reported in \cite{gen1, gen2, gen3}. CT suffers from some shortcomings. For instance, since the Cotton tensor vanishes identically when the Weyl tensor vanishes, this theory cannot be applied to non-trivial conformally flat spacetimes, like Friedmann-Lemaitre-Robertson-Walker (FLRW) metrics.  See  \cite{crit1, crit2, crit3, crit4, crit5, crit6} for some criticism on CT. In \cite{Gurses} a crucial distinction between CT and GR is reported; CT admits $pp$ waves with non-flat wave surfaces, in contrast to GR.  Also, AdS spherical and dS hyperbolic wave metrics in GR do not solve the field equations of CT. Quite recently, it has been shown that the CT has identically zero conserved charges for all of its solutions as long as there is no matter at the spatial boundary of spacetime \cite{Altas}. This represents that in CT the vacuum without a black hole is degenerate with a spacetime with many black holes.
 
 The original formulation of CT involves third rank tensorial equations with third-order derivatives, making them more challenging to solve than the field equations in GR. 
The field equations of CT are derived by requiring that both the second-order Schouten  $S_{\mu\nu}$ tensor and its associated source tensor $\mathcal{T}_{\mu\nu}$ (which is linearly related to the ordinary energy-momentum tensor $T_{\mu \nu}$ in GR) are Codazzi tensors, meaning they satisfy the Codazzi differential condition. Based on this fact, Mantica and Molinari introduced an alternative second rank formulation of CT called the Codazzi formulation \cite{Mantica}.
Based on this formulation,  a classification of vacuum and non-vacuum cases of field equations and their some non-trivial exact solutions were introduced by Sussman and
Najera \cite{gen3}. 

The Kiselev \cite{Kiselev} and Dymnikova \cite{Dymnikova} solutions are two non-vacuum solutions in GR, of particular interest due to their anisotropic sources and singularity structures. The Kiselev solution generalizes the Schwarzschild solution in the presence of surrounding fields, specifically the quintessence-like field which is often used to model dark energy. This addition to the standard Schwarzschild metric gives insights into how black holes might interact with dark energy or how black hole properties, such as their thermodynamics and event horizon structure might be affected in cosmologically relevant environments, see for instances \cite{yh1, yh2, yh3, yh4}. This generalization is well justified by the fact that real astrophysical black holes are neither isolated nor surrounded by empty space. Black hole solutions interacting with matter fields, like the Kiselev solution, are valuable for examining astrophysically distorted black holes \cite{bh1, bh2, bh3, bh4}  and for investigating the no-hair theorems \cite{bh5, bh6, bh7, bh8}. The Dymnikova solution provides a model of a regular black hole, where the singularity at the center is replaced by a regular de Sitter core. Hence, it is potentially offering a way to address one of the major issues in classical GR, the singularity problem. This solution motivated investigations for other regular black holes, see for instance \cite{reg1, reg2, reg3, reg4, reg5, reg6, reg7, reg8, reg9}. For the related works exploring these solutions in the context of theories beyond GR, see \cite{1a, 2a, 3a, 4a, 5a, 6a, 7a}.

In the present study, we aim to find non-vacuum solutions in CT akin to Kiselev and Dymnikova solutions in GR. 
Our motivation for investigating these solutions within CT is to reveal how this theory might account for or adapt to the effects of non-vacuum sources, and whether it can provide new insights into the behavior of both the singular and regular black holes in cosmological contexts. The organization of the paper is as follows. In Section 2, we give an introduction to CT and its Codazzi formulation. In Section 3, we find Kiselev and Dymnikova-type solutions in CT. The proofs for the derivation of these solutions are moved to Appendix I and II. Then, we address some aspects of the obtained solutions, in particular concerning singularities, thermodynamics, and geodesics, in comparison to the solutions in GR. Finally, in Section 4, we present our concluding remarks.

\section{Cotton Theory and Its Codazzi Formulation}
The original 3rd rank tensorial field equations of CT\footnote{We work in natural units, c=G=1.} were introduced by
Harada in \cite{Harada} as \begin{equation}
        C_{\nu\rho\sigma}= 16\pi \nabla_\mu T^{\mu}{}_{\nu\rho\sigma},\label{eq:GFE5}
\end{equation}
where $C_{\nu\rho\sigma}$ is the Cotton tensor and $T^{\mu\nu\rho\sigma}$ is the 4th rank tensor, called as generalized energy-momentum tensor, which is defined as 
\begin{eqnarray}
        T^{\mu\nu\rho\sigma}:=
        \frac{1}{2}(g^{\mu\rho}T^{\nu\sigma} - g^{\nu\rho}T^{\mu\sigma} - g^{\mu\sigma}T^{\nu\rho} + g^{\nu\sigma}T^{\mu\rho})-\frac{1}{6} (g^{\mu\rho}g^{\nu\sigma} - g^{\nu\rho}g^{\mu\sigma})T,
        \label{eq:T4tensor}
\end{eqnarray}
where $g^{\mu\nu}$ and $T= T_{\mu}^{\mu}$ are the spacetime metric and trace
of the ordinary energy-momentum tensor $T^{\mu\nu}$, respectively.
Soon later, Mantica and Molinari \cite{Mantica} showed that the field equations of CT can be written in a  Codazzi formulation as
\begin{eqnarray}
{ C}_{ab} &=& {\mathcal S}_{ab}-8\pi \mathcal{T}_{ab}, \nonumber \\
\nabla_aC_{bc}&=&\nabla_bC_{ac}\label{codazzi2}
 \end{eqnarray}
where $C_{ab}$ is the Codazzi tensor. Here, the Schouten tensor ${\mathcal S}_{ab}$  and the generalized energy-momentum tensor $\mathcal{T}_{ab}$ are 
\begin{eqnarray} {\mathcal S}_{ab} &=&  { R}_{ab}-\frac16{ R} g_{ab}= G_{ab}-\frac13 G g_{ab},\label{schouten} \nonumber \\
\mathcal{T}_{ab} &=& T_{ab}-\frac13 T g_{ab}.\label{Tgorda}
\end{eqnarray}A classification of vacuum and non-vacuum field equations of CT, based on the generalized energy-momentum tensor $\mathcal{T}_{ab}$, were introduced
by Sussman and Najera in \cite{gen3} as follows
\begin{itemize}
\item Vacuum: $\mathcal{T}_{ab}=T_{ab}-\frac13 T g_{ab}= 0$, hence
\begin{eqnarray}  
{\mathcal S}_{ab}={C}_{ab}\ne 0.\label{vac}
\end{eqnarray}
\item Non-vacuum: $\mathcal{T}_{ab}=T_{ab}-\frac13 T g_{ab}\ne 0$, hence
\begin{equation} {\mathcal S}_{ab}={C}_{ab}+8\pi\,\mathcal{T}_{ab},\quad {\mathcal S}_{ab}\ne \mathcal{C}_{ab},\label{nonvac}\end{equation}
\end{itemize}
In both the cases above, $C_{ab}$ satisfies the Codazzi condition \eqref{codazzi2}. As highlighted in \cite{gen3}, in the CT formulation, the Cotton tensor- based on the Weyl tensor- vanishes in conformally flat spacetimes, hindering meaningful non-vacuum solutions. The Codazzi formulation addresses this by keeping relevant tensors non-zero, enabling consistent application of CT field equations to a broader class of spacetimes, including non-vacuum and conformally flat ones \cite{m-mantica}. Moreover, while the CT formulation's covariant derivatives obscure the direct relationship between field sources and spacetime geometry, the Codazzi formulation presents the energy-momentum tensor directly in the equations. This facilitates clear physical interpretations and connects CT to GR results. Therefore, we are interested in Codazzi formulation to find non-vacuum solution of CT.
 \section{ Non-vacuum Solutions in Cotton Theory}
 In this section, we introduce two non-vacuum solutions of CT generalizing
 the known solutions in GR. 
\subsection{Generalized Kiselev Solution}

In the notable paper \cite{Kiselev}, Kiselev introduced new classes of static spherically symmetric exact solutions of the Einstein field equations with an anisotropic
fluid source. We give the following theorem for (generalized) Kiselev solution in CT. See Appendix I for the proof.
\\
\\
\noindent
\textbf{Theorem 1:}
CT admits Einstein GR's non-vacuum static spherically symmetric Kiselev solution. This theory generalizes the Kiselev solution for a single anisotropic fluid source to the solution possessing the line-element of the form
\begin{eqnarray}{\label{Kiselev general}}
    &&ds^2=-\Phi(r) dt^2+ \frac{dr^2}{\Phi(r)}+r^2 d\Omega^2,\nonumber\\
    &&\Phi(r)=1+\frac{c_0}{r}+c_1r+c_2r^2+c_3r^{-(1+3\omega)},
\end{eqnarray}
 and the energy-momentum tensor with the energy density  $\rho(r)$ as 
\begin{equation}
\rho(r)= 3\omega c_3 r^{-3 (1+\omega)},
\end{equation}
where $d\Omega^2=d\theta^2+\sin^2(\theta) d\phi^2$ represents the line-element of unit 2-sphere and  $c_0, c_1, c_2, c_3$ are arbitrary constants.
\\
 A particular case of the generalized Kiselev solution (\ref{Kiselev general}), was firstly obtained in \cite{Mantica2023}, where the authors introduced a new solution of CT in presence of a linear tensor of electrodynamics in Section 9.3.
Their metric function has the form
\begin{equation}
\Phi(r) = 1 - \frac{2M}{r} + \gamma r - \frac{\Lambda}{3}r^2 + \frac{q_e^2 + q_m^2}{r^2},
\end{equation}
where  $q_e$ and $q_m$ are the electric and magnetic charges, respectively. This represents a generalized Reissner-Nordstrom-(A)dS solution and corresponds to the specific case where $\omega = \frac{1}{3}$, $c_0=-2M$, $c_1=\gamma$, $c_2=-\frac{\Lambda}{3}$, and $c_3=q_e^2 + q_m^2$ in our analysis. It is important to note that Eq.(80) in \cite{Mantica2023} was obtained using a fully covariant approach to the original Harada's formulation of CT.
\\
\\
Here, we have the following immediate remarks.\\
\noindent
\textbf{Remark 1:} The vacuum solution of CT corresponds to $c_3=0$ \cite{Harada}. Hence
the nontrivial energy-momentum source induces the last term in 
\eqref{Kiselev general}.   \\
\textbf{Remark 2:} The new solution \eqref{Kiselev general}, generalizes
the Kiselev solution in GR with two new linear and quadratic terms.
In particular, the non-vacuum solutions in GR corresponding to quintessence field with $\omega=-2/3$ and cosmological constant with $\omega=-1$  emerge naturally as
vacuum solutions of CT.\\
\textbf{Remark 3:} The unique vacuum solution of GR, i.e. Schwarzschild solution, can be recovered in three cases: $(i)$ as a vacuum solution of the CT theory by
choosing $c_0=-2m, c_1=c_2=0$,  $(ii)$ as a non-vacuum solution of CT by choosing $c_0=-2m, c_1=-c_3, \omega=-2/3, c_2=0$, and $(iii)$ as a non-vacuum solution of CT by choosing $c_0=-2m, c_2=-c_3, \omega=-1, c_1=0$. Here $m$ is defined as the Schwarzschild mass.
\\
\textbf{Remark 4:} The case $\omega=1$ corresponds to a term $c_3/r^4$. Similar solutions can be found in \cite{r41, r42, heydar}.

\subsection{Analysis of the Solution}
In this section, we address some properties of the solution \eqref{Kiselev general} as follows.

\subsubsection{Arnowitt-Deser-Misner Mass}
For the case $c_1=c_2=c_3=0$, the metric \eqref{Kiselev general} is just the usual Schwarzschild metric and asymptotically flat. However, in the $c_1, c_2, c_3\neq0$ the metric is not asymptotically flat. The asymptotically flat boundary condition can be met for the following cases
\begin{equation}\label{asymphi}
  \Phi(r)\mid_{r\rightarrow\infty}=1 \left\{\begin{array}{l}
  \displaystyle  \mbox{for $c_1=c_2=c_3=0$}\\
  \\
  \displaystyle  \mbox{for $c_1=c_2=0,~~c_3\neq 0,~~\omega>-\frac{1}{3}$.}
  \end{array} \right.
\end{equation}

For stationary spacetimes possessing timelike Killing vector $K^\mu$, the Komar, defined in \cite{Komar}, and ADM masses are identical. The ADM mass is defined in \cite{ADM} as
\begin{equation}\label{adm}
M_{ADM}=-\frac{1}{4\pi}\int_{\mathcal{B}_\infty}\nabla^\mu K^\nu
d\Sigma_{\mu\nu},
\end{equation}
where $d\Sigma_{\mu\nu}=-u_{[\mu} v_{\nu]}dA$, with $dA=r^2\sin\theta d\theta d\phi$, is the differential surface element on a two-sphere $\mathcal{B}$ on a spacelike hypersurface $\Sigma$ of the spacetime. Here, $u_\mu=-\sqrt{\Phi}\delta^t_\mu$ and $v_\mu=\delta^r_\mu/\sqrt{\Phi}$ are the unit timelike and spacelike normals to $\mathcal{B}$, respectively, and $\mathcal{B}_\infty$ is a two-sphere at spatial infinity. Since the spacetime is stationary, it has the timelike Killing vector field $K^\mu=\delta^{\mu}_{t}$ and hence
\begin{equation}
\nabla^\mu K^\nu
d\Sigma_{\mu\nu}=-\frac{\Phi'}{2}dA,
\end{equation}
where the prime denotes differentiation with respect to $r$. Then, the ADM mass in (\ref{adm}) reduces to
\begin{equation}\label{admmass}
M_{ADM}=\frac{r^2}{2} \Phi' \mid_{r\rightarrow\infty}.
\end{equation}
Hence we find
\begin{equation}
M_{ADM}=\frac{1}{2}\left[ -c_0 + c_1 r^2 +2c_2 r^3 -c_3 (1+3\omega)r^{-3\omega} \right]\mid_{r\rightarrow
\infty}.
\end{equation}
To have an asymptotically well-defined ADM mass in CT, setting $c_0=-2m$, there are following three cases
\begin{equation}
M_{ADM} = m \begin{cases}
    \, \text{for } c_1 = c_2 = c_3 = 0, \\
    \\
    \, \text{for } c_1 = c_2 = 0,\ c_3 \neq 0,\ \omega = -\frac{1}{3}, \\
    \\
    \, \text{for } c_1 = c_2 = 0,\ c_3 \neq 0,\ \omega \geq 0.
\end{cases}
\end{equation}

\subsubsection{Essential Singularity, Black Hole Nature and Possible Horizons} The quadratic scalar of the Riemann curvature tensor, known as the Kretschmann scalar,  can be obtained for the generalized Kiselev solution \eqref{Kiselev general} as follows
\begin{eqnarray}\label{krshmn1}
 \mathcal{K}=R_{\alpha\beta\mu\nu} R^{\alpha\beta\mu\nu}&=& \frac{1}{r^6} \Big\{ \left[2 c_1+2 c_2 r^3+c_3 (9 \omega  (\omega +1)+2) r^{-3 \omega }\right]^2   + 4 r^{-6 \omega } \left[r^{3 \omega } \left(r^2 (c_0+c_2 r)+c_1\right)+c_3\right]^2 \nonumber\\
   && ~~~~~~+4 r^{-6 \omega } \left[r^{3 \omega } \left(c_1-r^2 (c_0+2 c_2 r)\right)+3 c_3 \omega +c_3\right]^2 \Big\}.
\end{eqnarray}
Hence, the solution has an essential singularity at $r=0$ as in GR. \\

 The possible event horizon(s) of the solution \eqref{Kiselev general} can be determined by the positive real roots of the equation $\Phi(r)=0$. This indeed depends on the sign and values of the parameters
$(c_0, c_1, c_2, c_3, \omega)$, and is not possible to determine in general.
An analysis can be done by looking at the metric function at the limits $\lim_{r\to 0^+}\Phi(r)$
and $\lim_{r\to \infty}\Phi(r)$. More explicitly, since $\Phi(r)$ is a continuous
function, there will be at least one real positive solution to $\Phi(r)=0$
if the limits have opposite signs. However, one notes that this is not the only case, and might be some other possibilities. Table \ref{table11} shows
some possibilities in which the solution \eqref{Kiselev general}  in CT possesses event horizon(s).
\begin{table}[!ht]
\centering
\begin{tabular}{|c|c|c|c|c|c|c|} 
\hline\hline
$w$ & $c_0$& $c_1$ & $c_2$ & $c_3$& $\lim_{r\rightarrow0^+}\Phi(r)$& $\lim_{r\rightarrow\infty}\Phi(r)$
\\ [0.5ex]
\hline 
$(-\infty,-5/3)$& $-$& $+$ &$\pm$& $+$ & $-$ &+\\[2ex]
$[-5/3,-1]$& $-$ &$\pm$ &$+$ &$\pm$& $-$ &$+$ \\[2ex]
$(-1, -2/3)$& $-$ &  $\pm$  & $\pm$ & $\mp$  & $\mp$ & $\pm$\\[2ex]
$(-2/3, \infty)$& $-$  &  $\pm$  & $+$ & $\pm$  & $-$ & $\pm$\\[2ex]
\hline\hline 
\end{tabular}
\caption{Some cases in which black holes certainly exist in CT.}\label{table11}
\end{table}
\\

Black hole solutions may have one or multiple horizons. Let $r=r_{0}$ be the largest root of $\Phi(r)=0$ and hence representing the black hole outermost event horizon.
In the case of a single event horizon, we can write the metric function $\Phi(r)$ as 
\begin{equation}\label{}
  \Phi(r)=(r-r_0)\psi(r),
\end{equation}
where $\psi(r)$ is a continuous function for $r\geq r_0$. One notes that here $\psi(r)>0$ since $\Phi(r)>0$ for $r>r_0$. Hence, we have
\begin{equation}\label{}
  \Phi'(r_0)=\psi(r_0)>0.
\end{equation}
In the case of multiple event horizons, the metric function $\Phi(r)$ can be written as
\begin{equation}\label{hmul}
  \Phi(r)=(r-r_1)(r-r_2)\ldots(r-r_m)\psi(r),
\end{equation}
where we assumed there are $m$ number of horizons, and $\psi(r)>0$ for $r$ greater than the largest root $r_0$, the outermost horizon radius. Again, due to the continuity of $\psi(r)$ for $r\geq r_0$,
\begin{equation}\label{}
  \Phi'(r_0)=\psi(r_0)>0,
\end{equation}
where we assumed that all the roots are distinct and the event horizon locates at $r_0$, the largest root of (\ref{hmul}).
When some or all of the roots are coincident, we have the extreme case. For example, for two coincident roots,
\begin{equation}\label{}
  \Phi(r)=(r-r_0)^2\psi(r),
\end{equation}
where $\psi(r)>0$ for $r>r_0$. Then
\begin{equation}\label{}
  \Phi'(r_0)=0,
\end{equation}
will be the indicator of an extreme black hole.

\subsubsection{Thermodynamics of Possible Black Holes}
Suppose $r_0$ locates the outermost black hole event horizon, that is $\Phi(r_0)=0$. Then, one can write the metric function (\ref{Kiselev general}) as 

\begin{equation}
\Phi(r)=1+\frac{c_{0}}{r}+\frac{c_{1} r_0^{3}}{r^{2}}\left(\frac{r}{r_{0}}\right)^{3}+c_{2} r^{2}+\frac{c_{3} r_{0}^{1-3\omega}}{r^{2}}\left(\frac{r}{r_{0}}\right)^{1-3\omega}.
\end{equation}
To realize thermodynamical characteristic of the parameters $c_0, c_1, c_2$ and $c_3$, one can define
\begin{equation}
c_{0}=-2m,~ Q_{1}=c_{1} r_{0}^{3},~ c_{2}=-\frac{\Lambda}{3},~ Q_{2}=c_{3} r_{0}^{1-3\omega},
\end{equation}
where $m, \Lambda, Q_1, \text{and } Q_2$ are the black hole mass, cosmological constant, and effective charges associated to CT and surrounding field, respectively. Hence, the metric function reads
\begin{equation}
\begin{aligned}
\Phi(r)= & 1-\frac{2 m}{r}-\frac{\Lambda}{3} r^{2}+\frac{Q_{1}}{r^{2}}\left(\frac{r}{r_{0}}\right)^{3}+\frac{Q_{2}}{r^{2}}\left(\frac{r}{r_{0}}\right)^{1-3\omega}.
\end{aligned}
\end{equation}
Hence, at $r=r_0$ it reduces to
\begin{equation}
\begin{aligned}
\Phi(r_0)= & 1-\frac{2 m}{r_{0}}-\frac{\Lambda}{3} r_{0}^{2}+\frac{Q_{1}}{r_0^{2}}+\frac{Q_{2}}{r_{0}^{2}}=0, \\
\end{aligned}
\end{equation}
and the horizon radius is the function $r_{0}=r_{0}\left(m, \Lambda, Q_{1}, Q_{2}\right).$ Considering the Hawking-Bekenstein black hole entropy (see \cite{Bekenstein} - \cite{Hawking}), $S=k A$, where $k$ is the Boltzman constant divided by Planck's length and $A= 4 \pi r_0^2$ is the area of the event horizon, one finds
\begin{equation}
\begin{aligned}
\delta S=k \delta A= 8 \pi k r_{0}\left(r_{0 m} \delta m+r_{0 \Lambda} \delta \Lambda+r_{01} \delta Q_{1}+ r_{02} \delta Q_{2}\right),
\end{aligned}
\end{equation}
where we defined $r_{0m}=\frac{\partial r_0}{\partial m},~r_{0\Lambda}=\frac{\partial r_0}{\partial \Lambda}, ~r_{01}=\frac{\partial r_{0}}{\partial Q_{1}}, ~r_{02}=\frac{\partial r_{0}}{\partial Q_{2}}$. 
Hence one finds the first law of black hole thermodynamics
\begin{equation}\label{flt_1}
\begin{aligned}
& \delta m=T \delta S+V \delta P+\Psi_{1} \delta Q_{1}+\Psi_{2} \delta Q_{2},
\end{aligned}
\end{equation}
where 
\begin{equation*}
    \begin{aligned}
        &T=  \frac{1}{8 \pi k r_0 r_0m} =\frac{1}{16 \pi k}\Phi^{'}(r_0),\\
        &V= 8 \pi \frac{r_{0 \Lambda}}{r_{0m}}=\frac{4 \pi}{3}r_0^3,\\
        &P= - \frac{\Lambda}{8 \pi},\\
        &\Psi_1= - \frac{r_{01}}{r_{0m}}= \frac{1}{2r_0},\\
        &\Psi_2=  - \frac{r_{02}}{r_{0m}}=\frac{1}{2r_0}.
    \end{aligned}
\end{equation*}
Here, $T, V, P ,\Psi_1, \text{and } \Psi_2$ are the black hole temperature, volume, pressure, and potentials associated to CT and the surrounding field, respectively.  Also, we have $\Phi^{'}(r_0)=\frac{2m}{r_0^2}-\frac{2 \Lambda}{3}r_0 - \frac{2(Q_1 +Q_2)}{r_0^3}+\frac{1}{r_0^2}\left(\frac{d(Q_1+Q_2)}{dr_0}\right)$. Here, the derivative of the metric function \(\Phi^{'}(r_0)\) differs from the Reissner-Nordstrom-de Sitter black hole solution in GR by the term \(\frac{1}{r_0^2}\left(\frac{d(Q_1 + Q_2)}{dr_0}\right)\) due to the effective charges associated to CT (via $c_1$ parameter) and surrounding field (via $c_3$ parameter). Since the Hawking temperature \(T\) is proportional to the surface gravity \(\kappa\), the surface gravity of the generalized Kiselev black holes in CT also differs from that in GR unless \(\frac{1}{r_0^2}\left(\frac{d(Q_1 + Q_2)}{dr_0}\right) = 0\).
 Moreover, by defining the total charge $Q_t= Q_1 + Q_2$ and potential $\Psi=\Psi_1=\Psi_2$, one can write \eqref{flt_1} as
\begin{equation}
    \delta{m}=T \delta S+V \delta P+\Psi \delta Q_{t}.
\end{equation}

\subsubsection{Null and Timelike Geodesics}
The metric has two killing vectors 
$ k^{\mu}=\left(\partial_{t}\right)^{\mu}=(1,0,0,0)$  and $l^{\mu}=\left(\partial_{\phi}\right)^{\mu}=(0,0,0,1)$. 
The corresponding conserved quantities are the total energy $E$ and  angular momentum $L$, respectively, as
\begin{eqnarray}
&& E=-k_{\mu} \frac{d x^{\mu}}{d \lambda}=\Phi\left(\frac{d t}{d \lambda}\right),\nonumber \\
&& L=l_{\mu} \frac{d x^{\mu}}{d \lambda}=r^{2}\left(\frac{d \phi}{d \lambda}\right),
\end{eqnarray}
 where $\lambda$ is an affine parameter along the geodesics, and $\Phi$ is given in \eqref{Kiselev general}. The geodesic equation $ g_{\mu \nu} \frac{d x^{\mu}}{d \lambda} \frac{d x^{\nu}}{d \lambda}=\epsilon$ with $\epsilon= 1, 0, -1  $ for the spacelike, null and timelike geodesics, respectively, takes the form
\begin{equation}
-\Phi^{2}\left(\frac{d t}{d \lambda}\right)^{2}+\left(\frac{d r}{d \lambda}\right)^{2}+\Phi\left[r^{2}\left(\frac{d \phi}{d \lambda}\right)-\epsilon\right]=0.
\end{equation}
Defining
\begin{eqnarray}{\label{geodesic 2}}
 \varepsilon&=&E^{2} / 2, \nonumber\\
 V_{eff}&=&\frac{1}{2} \Phi\left(\frac{L^{2}}{r^{2}}-\epsilon\right),
\end{eqnarray}
the geodesic equation reads as
\begin{equation}{\label{geodesic 1}}
\frac{1}{2}\left(\frac{d r}{d \lambda}\right)^{2}+V_{eff}=\varepsilon,
\end{equation}
where the effective potential has the following explicit form
\begin{eqnarray}
V_{eff}&= & \frac{1}{2}\left[1-\frac{2 m}{r}+c_2 r^{2}+c_{1} r+c_{3} r^{-(1+3 \omega)}\right]\left(\frac{L^{2}}{r^{2}}-\epsilon\right) \nonumber\\
&= & -\frac{\epsilon}{2}+\frac{\epsilon m}{r}+\frac{L^{2}}{2 r^{2}}-\frac{m L^{2}}{r^{3}}
 +\frac{c_{1} L^{2}}{2 r}-\frac{\epsilon c_{1}}{2} r + \frac{c_2 L^2}{2}-\frac{\epsilon c_2}{2}r^2
 +\frac{c_{3} L^{2}}{2} r^{-3(1+\omega)}-\frac{\epsilon c_{3}}{2} r^{-(1+3 \omega)}.
\end{eqnarray}
The first four terms are standard GR terms, but the remaining are new correction terms induced by the cosmological constant, CT and surrounding field.

Circular orbits at radii $r_{c}$ occur where the effective potential $V_{eff}$ is flat, meaning that $\left.\frac{d V_{eff}}{d r}\right|_{r=r_{c}}=0$.
Hence, one obtains the equation for circular orbits as
\begin{equation}{\label{cicular orbit eq}}
\begin{aligned}
- & \frac{\epsilon m}{r_{c}^{2}}-\frac{L^{2}}{r_{c}^{3}}+\frac{3 m L^{2}}{r_{c}^{4}}
 -\frac{c_{1}}{2}\left(\epsilon+\frac{L^{2}}{r_{c}^{2}}\right)-\epsilon c_2 r 
+\frac{c_{3}}{2}\left[-3(1+\omega) L^{2} r_{c}^{-(3 \omega+4)}+\epsilon(1+3 \omega) r_{c}^{-(3 \omega+2)}\right]=0.
\end{aligned}
\end{equation}
We have the following remarks regarding the circular orbits in CT.\\
\noindent
\textbf{Remark 5:} GR limit can be recovered, by setting
 $c_{1}=c_{2}=c_{3}=0$, as
\begin{equation}
 \quad-\frac{\epsilon m}{r_{c}^{2}}-\frac{L^{2}}{r_{c}^{3}}+\frac{3 m L^{2}}{r_{c}{ }^{4}}=0,
\end{equation} 
or equivalently,
\begin{equation}
\epsilon m r_{c}^{2}+L^{2} r_{c}-3 m L^{2}=0.
\end{equation}
Solving this equation yields the circular radii $r_c$ as follows
\begin{equation}
\begin{cases}r_{c}=3 m & \epsilon=0 \\
\\r_{c}^{ \pm}=\frac{L^{2} \pm \sqrt{L^{4}-12 m^{2} L^{2}}}{2 m} & \epsilon=-1.\end{cases}
\end{equation}
\textbf{Remark 6:} The parameter $c_2$ in CT, which can be interpreted as cosmological constant, does not contribute to null geodesics, since $\epsilon=0$.
However, $c_1$ and $c_3$ parameters, characterizing the CT and surrounding field, affect both the null and timelike geodesics.

\noindent\textbf{Remark 7:} For null geodesics, there exists a particular orbit $r_{c}=3m$ provided that 
\begin{equation*}
\frac{c_1}{c_3}= - (1+ \omega) 3^{-(3 \omega +1)} m^{-(3\omega + 2)},
\end{equation*}
such that the linear contribution of the CT via the $c_1$ parameter does not appear.

\noindent\textbf{Remark 8:} Null geodesics are given by the solutions to
\begin{equation}{\label{Eq. square}}
{c_{1}}  r_{c}^{2}+2r_{c}-6 m +{3 c_{3}(1+\omega) } r_{c}^{-3 \omega}=0.
\end{equation}
One can consider the following interesting cases.
\begin{itemize}
    \item [(i)] For $\omega=0$, describing the dust-like surrounding field, Eq.\eqref{Eq. square} takes the form
\begin{equation}
 {c_{1}}  r_{c}^{2}  + 2r_{c}-3 \left(2m-{c_{3}}\right)=0.
\end{equation}
Then, there are solutions 
\begin{equation}
\begin{cases}r_{c}=\frac{6m-3c_3}{2} & {c_1} = {0} \\
\\
r_{c}^{ \pm}=\frac{ -1 \pm \sqrt{1+ 3c_1\left( 2m-{c_{3}} \right)}}{c_1} & c_1\neq 0,\end{cases}
\end{equation}
where $-\frac{1}{3\left(2m -{c_3}\right)}\leq c_1<0$ for $r_c^-$.
    \item[(ii)] For $\omega=-1$, describing the cosmological constant-like surrounding field, Eq.\eqref{Eq. square} takes the form

    \begin{equation}
       {c_{1}}  r_{c}^{2}  +2r_{c}-6 m =0,
    \end{equation}
where the effect of the surrounding field via its parameter $c_3$ does not appear. Then, there are solutions
    \begin{equation}
\begin{cases}r_{c}=3m & {c_1} = {0} \\
\\
r_{c}^{ \pm}=\frac{ -1 \pm \sqrt{1+6 mc_1}}{c_1} & c_1 \neq 0,\end{cases}
\end{equation}
where $-\frac{1}{6m}\leq c_1<0$ for $r_c^-$. Here, the first solution coincides with the GR limit for $\epsilon=0$.
    \item[(iii)] For $w=-2 / 3$, describing an exotic surrounding field, Eq.(\ref{Eq. square}) takes the form
    \begin{equation}
\left( {c_1}+{c_3}\right) r_c^2 +2r_c -6m =0.
    \end{equation}
Then, there are solutions 
\begin{equation}
\begin{cases}r_{c}=3m & c_1=-c_3 \\
\\
r_{c}^{ \pm}=\frac{ -1 \pm \sqrt{1+ 6m\left(c_1 + c_3 \right)}}{c_1+ c_3} & c_1\neq -c_3,\end{cases}
\end{equation}
where $-\frac{1}{6m}\leq c_1+c_3<0$ for $r_c^-$. Here, the first solution coincides with the GR limit for $\epsilon=0$.
\end{itemize}
\textbf{Remark 9:} From \eqref{cicular orbit eq}, the equation governing the timelike geodesics is 
\begin{equation}{\label{Eq. star}}
 m r_{c}^{2}-L^{2} r_{c}+3 m L^{2} -\frac{c_{1}}{2}\left(\frac{L^{2}}{r_{c}^{2}}-1\right) r_{c}^{4}+c_2 r_c^5 -\frac{c_{3}}{2}\left[3(1+\omega) L^{2} r_{c}^{-3 \omega}+(1+3 \omega) r_{c}^{-3 \omega+2}\right]=0.
\end{equation}
Solving this equation for generic $\omega$ is not possible. Hence, to realize the effect of CT via its linear term parameter $c_1$ and surrounding field parameter $c_3$, one can set $c_2=0$ and consider the following cases.
\begin{itemize}
    \item [] {\bf (1) For large $r$ values}: 
    \begin{itemize}
        \item[(i)] Setting $c_2=0, \omega>0$, Eq.\eqref{Eq. star} takes the form
        \begin{equation}
             c_1 r_c^4 + (2m- c_1 L^2)r_c^2\approx0.
        \end{equation}
        When $c_1 =0$, $r_c \approx 0$. Then, when $c_1 \neq 0$, there are two approximate solutions
        \begin{eqnarray}
            &{r_c}^+\approx \sqrt{L^2-\frac{2m}{c_1}}&, \nonumber\\
            &{r_c}^-\approx 0&.
        \end{eqnarray}
        \item[(ii)] Setting $c_2=0, \omega=0$, Eq.\eqref{Eq. star} takes the form
        \begin{equation}
            {c_1} r_c^4 + (2 m- {c_1}L^2- c_3 )r_c^2 \approx0.
        \end{equation}
       When $c_1 =0$ and $c_1 L^2 + c_3 \neq 2m$, $r_c \approx 0$. Then, when $c_1 \neq 0$, there are two approximate solutions
        \begin{eqnarray}
            &{r_c}^+\approx \sqrt{L^2-\frac{2m-c_3}{c_1}}&, \nonumber\\
            &{r_c}^-\approx 0&.
        \end{eqnarray}
        \item[(iii)] Setting $c_2=0, \omega=-\frac{2}{3}$, Eq.\eqref{Eq. star} takes the form
        \begin{equation}
            \left(c_1 + c_3\right) r_c^4 + \left( 2m - (c_1 + c_3) L^2 \right)r_c^2 \approx 0.
        \end{equation}
        When $c_1 = -c_3$ and $c_1 + c_3 \neq \frac{2m}{L^2}$, $r_c \approx 0$. When, $c_1 \neq -c_3$,  there are two approximate solutions
        \begin{eqnarray}
            &r_c^+\approx \sqrt{L^2-\frac{2m}{c_1 + c_3}}&,\nonumber\\
            &r_c^-\approx 0&.
        \end{eqnarray}
    \end{itemize}

    \item[] {\bf (2) For small $r$ values}: Neglecting the higher order terms, one finds the following cases.  
    \begin{itemize}
        \item [(i)] For $ \omega<-\frac{2}{3}$, Eq.\eqref{Eq. star} takes the form
        \begin{equation}
            (2m- c_1 L^2) r_c^2 - 2 L^2 r_c + 6mL^2\approx 0.
        \end{equation}
        There are approximate solutions
        \begin{equation}
        \begin{cases}
        r_c\approx 3m & c_1= \frac{2m}{L^2}\\
        \\
            r_c^{\pm}\approx \frac{L^2 \pm \sqrt{L^4 - 6mL^2(2m-c_1 L^2)}}{2m-c_1 L^2} & c_1 \neq \frac{2m}{L^2}.
        \end{cases}
        \end{equation}
        Here, the first case with $\epsilon=-1$ almost coincides with the GR limit with $\epsilon=0$.
        \item [(ii)] For $\omega=0$, Eq.\eqref{Eq. star} takes the form
        \begin{equation}
            (2m- c_1 L^2 - c_3)r_c^2 - 2L^2 r_c + (6m-c_3)L^2 \approx 0.
        \end{equation}
        There are approximate solutions
        \begin{equation}
        \begin{cases}
            r_c \approx \frac{6m - c_3}{2} & c_1 L^2 + c_3 = 2m\\
            \\
            r_c^{\pm}\approx\frac{L^2 \pm \sqrt{L^4 - L^2(6m - c_3)(2m-c_1 L^2 -c_3)} }{(2m-c_1 L^2 -c_3)} & c_1 L^2 + c_3 \neq 2m.
            \end{cases}
        \end{equation}
        \item[(iii)] For $\omega=-\frac{2}{3}$, Eq.\eqref{Eq. star} takes the form
        \begin{equation}
            \left( 2m -(c_1 + c_3)L^2 \right) r_c^2 - 2 L^2 r_c + 6mL^2\approx 0.
        \end{equation}
        There are approximate solutions
        \begin{equation}
            \begin{cases}
                r_c\approx 3m & c_1 + c_3 = \frac{2m}{L^2}\\
                \\
                r_c^{\pm}\approx \frac{L^2 \pm \sqrt{L^4 - 6mL^2(2m -  (c_1 + c_3)L^2)}}{2m - (c_1 + c_3)L^2} & c_1 +c_3 \neq \frac{2m}{L^2}.
            \end{cases}
        \end{equation}
         Here, the first case with $\epsilon=-1$ almost coincides with the GR limit with $\epsilon=0$.
        \item [(iv)] For $ \omega=\frac{1}{3}$, Eq.\eqref{Eq. star} takes the form
        \begin{equation}
            r_c^2 - 6m r_c + 2c_3  \approx 0.
        \end{equation}
        There are approximate solutions
        \begin{equation}
            r_c^{\pm}\approx {3m \pm \sqrt{9m^2 - 2 c_3}}.
        \end{equation}
  Here, both cases when $c_3= \frac{9m^2}{2}$ with $\epsilon=-1$ almost coincides with the GR limit with $\epsilon=0$.
    \end{itemize}
\end{itemize}


\subsection{Generalized Dymnikova Solution}
Another notable static spherically symmetric non-vacuum solution in GR was introduced by Dymnikova \cite{Dymnikova}. This solution represents a non-singular black hole solution where instead of a point-like singularity, this solution possesses a de Sitter-like core at the center of the black hole. Hence, the metric describes a transition from (an internal) de Sitter space to (an external) Schwarzschild-like solution.
This solution characterizes a black hole with two horizons, in which the finiteness of all curvature invariants confirms
its non-singular nature. We give the following theorem for (generalized) Dymnikova solution in CT. See Appendix II for the proof.
\\
\\
\noindent
\textbf{Theorem 2:} CT admits Einstein GR's non-vacuum static spherically symmetric Dymnikova solution. This theory generalizes the Dymnikova solution for a single anisotropic fluid source to the solution possessing the line-element of the form
\begin{eqnarray}\label{metdym_1}
    &&ds^2=-\Phi(r) dt^2+ \frac{dr^2}{\Phi(r)}+r^2 d\Omega^2,\nonumber\\
&&\Phi(r) = 1 +  \frac{c_0}{r}\left( 1-\exp\left(-\frac{r^3}{r_*^3}\right) \right) + c_1 r  + c_2 r^2\nonumber\\
&&~~~~~~~=1 -  \frac{R_g(r)}{r} + c_1 r  + c_2 r^2,
\end{eqnarray}
 with the energy-momentum tensor having the energy density  $\rho(r)$ as 
\begin{eqnarray}{\label{-rho}}
    \rho(r)=\rho_0\exp\left(-\frac{r^3}{r_*^3 }\right),
\end{eqnarray}
where $d\Omega^2=d\theta^2+\sin^2(\theta) d\phi^2$ represents the line-element of unit 2-sphere, and $c_0, c_1, c_2, \rho_0$ are arbitrary constants.
 Here, we defined $R_g(r)=-c_0\left( 1-\exp\left(-\frac{r^3}{r_*^3}\right) \right)$ with $r_*^3=-c_0 r_0^2 $ where $r_0$ represents the core de Sitter radius $r_0=\sqrt{ 3/\rho_0}$. Regarding the metric, we have the following immediate remark.\\
\\
{\bf Remark 10:} In contrast to the original Dymnikova solution in GR, the generalized solution with the metric function \eqref{metdym_1} in CT (with the same single source in GR having the energy density \eqref{-rho}) possesses two distinct de Sitter regions: the de Sitter core (related to $\rho_0$) and the de Sitter asymptote (related to $c_2$).

Here, it is worth mentioning that the $c_2$ parameter, characterizing the outer de Sitter region,   in the solutions of CT has a pure geometric
origin. This parameter is not related to the source of field equations and
it appears even in vacuum solutions of CT \cite{Harada}. 
In contrast, in GR, the induced (anti) de Sitter term in any metric comes
from including $\Lambda g_{\mu\nu}$ as a source of Einstein field equations. However,
 solutions of CT naturally have this term, and  {\it $c_2$ as the cosmological constant}, appears naturally as {\it an integration constant}. Although $c_2$ in CT plays the role of cosmological constant in GR, however, their origins are different.
This signifies an interesting distinction between CT and GR.
 The existence of the core de Sitter region parametrized by  $\rho_0$ is because of the inclusion
of the nontrivial source with the energy density (61). Here, one should note that this source is asymptotically vanishing, i.e. $\rho \to 0$ as $r\to \infty$. Hence this source does not exist/dominate at the asymptotic region of spacetime and cannot play the role of the cosmological constant. Hence, $c_2$ (as the cosmological constant) and $\rho_0$ parameters have different origins and exist independently in CT.
 This fact is supported by the result of Mantica and Molinari in \cite{Mantica}, where they also investigated CT in constant curvature $n$-dimensional spacetimes. Specifically, in Section 3.6, they pointed out that the only non-trivial Codazzi tensor in a space of constant curvature is given by
\begin{equation}
C_{ab}=\nabla_{a} \nabla_{b} \phi+\frac{R \phi}{n(n-1)} g_{ab},
\end{equation}
where $\phi$ is an arbitrary scalar function. This result implies that de Sitter, Minkowski, and anti-de Sitter spacetimes naturally emerge as non-vacuum solutions of CT with a non-trivial stress-energy tensor given by
\begin{equation}
T_{ab}=g_{ab}\left[\nabla_{c} \nabla^{c} \phi+\frac{R}{n} \phi-\frac{n-2}{2n} R\right]+\nabla_{a} \nabla_{b} \phi.
\end{equation}
Vacuum solutions are obtained if and only if $\phi$ satisfies
\begin{equation}
\nabla_{a} \nabla_{b} \phi=g_{ab}\left[-\frac{R}{n(n-1)} \phi+\frac{n-2}{2 n(n-1)} R\right],
\end{equation}
which leads to the trivialization of the Codazzi tensor as
\begin{equation}
C_{ab}=\frac{R(n-2)}{2 n(n-1)} g_{ab}.
\end{equation}

\subsection{Analysis of the Solution}
In this section, we address some properties of the generalized Dymnikova solution \eqref{metdym_1} as follows.
\subsubsection{Arnowitt-Deser-Misner Mass}
Regarding the metric function \eqref{metdym_1}, one obtains the ADM mass using \eqref{admmass} as 
\begin{equation}
\begin{aligned}
 M_{A D M} & = \frac{1}{2}\left[-c_0\left(1-\exp \left(-r^3 / r_{*}^3\right)\right)+\frac{3 c_0 r^3}{r_{* }^3} \operatorname{exp}\left(-r^3 / r_{* }^3\right)+c_1 r^2+2 c_2 r^3\right]\mid_{r\rightarrow
\infty} \\
& =\frac{1}{2}\left[-c_0+ c_1 r^2+2 c_2 r^3\right]\mid_{r\rightarrow
\infty}.
\end{aligned}
\end{equation}
 Hence, an asymptotically well-defined ADM mass for the generalized Dymnikova solution in CT is provided by setting $c_1=c_2=0$ as
\begin{equation}
M_{A D M}=-\frac{1}{2} c_0=m,
\end{equation}
where we have defined $c_0=-2 m$.

\subsubsection{Essential Singularity, Black Hole Nature and Possible Horizons}
The Kretschmann scalar for the generalized Dymnikova solution \eqref{metdym_1} takes the following form
\begin{equation}
    \label{krshmn2}
\mathcal{K}=4\left(\frac{R_g(r)}{r^3}-\frac{c_1 +c_2 r}{r}  \right)^2  +4\left(\frac{R_g(r)}{r^3}-\frac{R_g^\prime(r)}{r^2}+\frac{c_1 +2c_2 r}{r}  \right)^2
+\left(\frac{2R_g(r)}{r^3}-\frac{2R_g^\prime(r)}{r^2}+\frac{R_g^{\prime\prime}(r)}{r} -2c_2   \right)^2.
\end{equation}
\textbf{Remark 11:} In contrast to GR, the generalized solution with the metric function \eqref{metdym_1} in CT has a diverging Kretschmann scalar as $r\to 0$. In fact, the solution \eqref{metdym_1} in CT is singular due to the linear term, i.e. $c_1 r$, resulting from the higher derivative nature of the theory. If $c_1=c_2=0$, the general solution \eqref{metdym_1} reduces to the original Dymnikova solution with a finite Kretschmann scalar at the center and remains non-singular everywhere. The general solution \eqref{metdym_1} still remains non-singular for $c_1=0$ and $c_2\neq 0$.

The solution \eqref{metdym_1} can represent a black hole since there will be at least one real positive solution to $\Phi(r)=0$. Table \ref{table22} shows a possibility in which the solution \eqref{metdym_1} in CT possesses event horizon(s).
\begin{table}[!ht]
\centering
\begin{tabular}{|c|c|c|c|c|} 
\hline\hline
 $c_0$& $c_1$ & $c_2$ & $\lim_{r\rightarrow0^+}\Phi(r)$& $\lim_{r\rightarrow\infty}\Phi(r)$
\\ [0.5ex]
\hline 
 $-$& $\pm$ &$-$&  $+$ &$-$\\[2ex]
\hline\hline 
\end{tabular}
\caption{Two cases in which black holes certainly exist in CT.}\label{table22}
\end{table}

Finding the exact location of horizons for \eqref{metdym_1} is not possible. Hence, following Dymnikova \cite{Dymnikova}, we determine approximate horizon locations. We do this approximation separately for cosmological and astrophysical scales. The first locates the cosmological horizon(s) $R_c$ and the latter gives de Sitter core and black hole event horizon radii much smaller than $R_c$. 

For cosmological scales, we have 
\begin{equation}
    \begin{aligned}
        \Phi(r) \approx 1+ c_1 r+ c_2 r^2.
    \end{aligned}
\end{equation}
This gives the cosmological horizon location as 
\begin{equation}
    \begin{aligned}
        R^{\pm}_c\approx \frac{-c_1}{2c_2} \pm \sqrt{\frac{c_1^2}{4c_2}-\frac{1}{c_2}}. 
    \end{aligned}
\end{equation}
When $c_1=0$, one finds the usual de Sitter radius $\sqrt{-1/c_2}$ with $c_2<0$. Here, the linear contribution $c_1 r$ of CT affects the size of the cosmological horizon. 

For astrophysical scales, i.e  $r \ll R_c$,  we have 

\begin{equation}
    \begin{aligned}
        \Phi(r) \approx 1 +  \frac{c_0}{r}\left( 1-\exp\left(-\frac{r^3}{r_*^3}\right) \right),
    \end{aligned}
\end{equation}
Here, by the Maclurin series expansion, one finds
\begin{eqnarray}
        &&r^{+}\approx -c_0\left( 1-\mathcal{O}\left(\exp\left(-\frac{c_0^2}{r_0^2}\right)\right)\right),\nonumber \\
        &&r^{-} \approx r_0\left( 1-\mathcal{O}\left(\exp\left(\frac{r_0}{2c_0}\right)\right)\right).
\end{eqnarray}
These represent the modified Schwarzschild and the core de Sitter radii, respectively. 
Here one observes an effective (smaller) Schwarzschild mass $M_{eff}=2m\left( 1-\mathcal{O}\left(\exp\left(-\frac{4 m^2}{r_0^2}\right)\right)\right)$ setting $c_0=-2m$.

 \subsubsection{Thermodynamics of Possible Black Holes}
When $r \ll R_{c}$ one can write the metric function Eq.(\ref{metdym_1}) as 
\begin{equation}
\begin{aligned}
\Phi(r)\approx 1- \frac{2m}{r}\left(1- \exp \left( \frac{r^3}{2m r_0^2}\right)\right).
\end{aligned}
\end{equation}
by defining the parameter $c_0=-2m$. Hence, at $r=r^{+}$ it reduces to
\begin{equation}\label{Dymnikova outer}
\begin{aligned}
\Phi(r^+)\approx & 1-\frac{M_{eff}}{r^{+}}\approx 0, \\
\end{aligned}
\end{equation}
where $M_{eff}=2m\left( 1-\mathcal{O}\left(\exp\left(-\frac{4 m^2}{r_0^2}\right)\right)\right)$ and the horizon radius is the function $r^{+}=r^{+}(m).$ Considering the black hole entropy, $S=k A$, where $k$ is the Boltzman constant divided by Planck's length and $A= 4 \pi (r^{+})^2$ is the area of the event horizon, we have
\begin{equation}
\begin{aligned}
\delta S=k \delta A= 8 \pi k r^{+}r_{m}^+ \delta m,
\end{aligned}
\end{equation}
where we defined $r_{m}^+=\frac{\partial r^{+}}{\partial m}$. 
Hence, one finds the first law of black hole thermodynamics
\begin{equation}\label{flt}
\begin{aligned}
& \delta m=T \delta S,
\end{aligned}
\end{equation}
where 
\begin{equation}
    \begin{aligned}
        &T=  \frac{1}{8 \pi k r^+ r^{+}_m}\approx \frac{1}{16 \pi k}\frac{d \Phi(r^{+})}{d r^{+}},
    \end{aligned}
\end{equation}
is the temperature of the event horizon $r^{+}.$
\subsubsection{Null and Timelike Geodesics}
Using Eq.\eqref{Dymnikova outer}, one can find the effective potential for $r \ll R_c$ limit 
\begin{equation}
    V_{eff} \approx \frac{1}{2}\left( 1-\frac{M_{eff}}{r}\right)\left( \frac{L^2}{r^2} - \epsilon \right)=-\frac{\epsilon}{2}+\frac{\epsilon M_{eff}}{r}+\frac{L^{2}}{2 r^{2}}-\frac{M_{eff} L^{2}}{r^{3}},
\end{equation}
where $M_{eff}=2m\left( 1-\mathcal{O}\left(\exp\left(-\frac{4 m^2}{r_0^2}\right)\right)\right)$ is the effective Schwarzschild mass. 

Circular orbits $r_{c}$ are given by radii where the effective potential is flat, that is $\left.\frac{d V_{eff}}{d r}\right|_{r=r_{c}}=0$. 
Hence, one obtains the equation for circular orbits as
\begin{equation}
\begin{aligned}
- & \frac{\epsilon M_{eff}}{r_{c}^{2}}-\frac{L^{2}}{r_{c}^{3}}+\frac{3 M_{eff} L^{2}}{r_{c}^{4}} \approx 0,
\end{aligned}
\end{equation}
or equivalently,
\begin{equation}
\epsilon M_{eff} r_{c}^{2}+L^{2} r_{c}-3 M_{eff} L^{2} \approx 0.
\end{equation}
This equation has the following two solutions
\begin{equation}
\begin{cases}r_{c} \approx 3 M_{eff} & \epsilon=0 \\
\\r_{c}^{ \pm} \approx \frac{L^{2} \pm \sqrt{L^{4}-12 M_{eff}^{2} L^{2}}}{2 M_{eff}} & \epsilon=-1.\end{cases}
\end{equation}
\section{Conclusion}
Recently, critical distinctions between CT, a theory of third-rank tensorial field equations, and GR concerning wave and vacuum solutions have been addressed. The present study aims to reveal how CT might account for or adapt to the effects of non-vacuum sources. In particular, it examines whether CT can offer novel insights into the behavior of singular and regular black holes in astrophysical contexts, in comparison to GR. It is shown that CT
generalizes two particular non-vacuum black hole solutions of GR: Kiselev and Dymnikova solutions. Some aspects of these generalized solutions are as follows. 

The generalized Kiselev solution in CT extends the original solution in GR, which describes a black hole surrounded by a single anisotropic fluid source, by introducing two additional terms of geometric origin in the metric function: a linear term $c_1 r$ and a quadratic term $c_2 r^2$. Hence the parameter space of the generalized Kiselev solution in CT can be denoted by $(c_0, c_1, c_2, c_3, \omega)$ where they represent the Schwarzscild mass, linear and quadratic terms, the surrounding field and its equation of state parameters, respectively. 
Here are our remarks regarding the generalized Kiselev solution:
\begin{itemize}
\item The Kiselev solution in GR corresponding to the quintessence field $(\omega=-\frac{2}{3})$ and cosmological constant $(\omega= -1)$ emerge naturally as vacuum solutions of CT.
\item The Schwarzschild solution in CT, can be recovered in three cases: $(i)$ as a vacuum solution of the CT theory by
choosing $c_0=-2m, c_1=c_2=0$,  $(ii)$ as a non-vacuum solution of CT by choosing $c_0=-2m, c_1=-c_3, \omega=-\frac{2}{3}, c_2=0$, and $(iii)$ as a non-vacuum solution of CT by choosing $c_0=-2m, c_2=-c_3, \omega=-1, c_1=0$. Here $m$ is defined as the Schwarzschild mass.
\item The ADM mass will be asymptotically well-defined and match the
Schwarzschild black hole solution in GR in three cases: $(i)~ c_1 = c_2 =  c_3=0, (ii)~c_1 = c_2 = 0,\ c_3 \neq 0,\ \omega = -\frac{1}{3}$, and $(iii)~ c_1 = c_2 = 0,\ c_3 \neq 0,\ \omega \geq 0$.
\item The generalized Kiselev solution suffers from an essential singularity at $r = 0$, as in GR.
\item This solution may possess multiple distinct or extremal horizons depending on its parameters  $(c_0, c_1, c_2, c_3, \omega)$. The thermodynamics of an event horizon can be formulated akin to a Reissner-Nordstrom-de Sitter black hole in GR by introducing virtual charges associated with the  $c_1$ and $c_3$ parameters. The presence of virtual charges generalizes the first law of black hole thermodynamics with a modified Hawking temperature and two charge-induced potential terms.
\item  The parameter $c_2$ in CT does not contribute to null geodesics.
However, $c_1$ and $c_3$ parameters, affect both the null and timelike geodesics. Moreover, for null geodesics, the condition $ \frac{c_1}{c_3}= - (1+ \omega) 3^{-(3 \omega +1)} m^{-(3\omega + 2)}$ provides specific cases in which the circular orbits $r_c$ coincide exactly with the one in GR, i.e. $r_c=3m$. For instance, this occurs when $\omega=-\frac{2}{3},~ c_1=-c_3$. In such cases, CT and GR cannot be distinguished from the perspective of null circular orbits.
Similarly, there are also particular cases where the timelike geodesics almost coincide with $r_c=3m$ in GR for null geodesics. For instance, in small radial distances $r$ limit,  this occurs for $\omega < -\frac{2}{3}, c_1 = \frac{2m}{L^2}$, or for $\omega = -\frac{2}{3}, c_1 + c_3 = 2m$.
\end{itemize}

Similarly, CT generalizes the Dymnikova solution in GR, which describes a regular black hole with an anisotropic fluid source, by introducing the same additional terms $c_1 r$ and $c_2 r^2$. Hence the parameter space of the generalized Dymnikova solution reads as $(c_0, c_1, c_2, \rho_0)$ where they represent the Schwarzscild mass, linear and quadratic terms of geometric origin, and the core de Sitter energy density parameter, respectively. Some properties of the generalized Dymnikova solution are as follows.
\begin{itemize}
\item Unlike the original Dymnikova solution in GR, the generalized solution in CT possesses two distinct de Sitter regions: the de Sitter core (related to $\rho_0$) and the de Sitter asymptote (related to $c_2$), having the source and geometric origins, respectively.
\item The ADM mass is asymptotically well-defined and matches the
Schwarzschild case in GR only if $ c_1 = c_2 =0$.
\item  In contrast to GR, the generalized Dymnikova solution in CT has a diverging Kretschmann scalar at $r= 0$. This solution in CT is singular due to the linear term, i.e. $c_1 r$, resulting from the higher derivative nature of the theory. If $c_1=c_2=0$, the generalized solution here reduces to the original Dymnikova solution with a finite Kretschmann scalar at the center and remains regular everywhere. The generalized Dymnikova solution also remains regular for $c_1=0$ and $c_2\neq 0$.
\item  The thermodynamics and geodesic structure are akin to the Schwarzschild case in GR. However, the effective mass here is smaller than the Schwarzschild mass in GR.
\end{itemize}

\section*{Acknowledgment}
M. Ranjbar would like to express her sincere gratitude to Bilkent University for their generous hospitality during the summer of 2024. The university's supportive environment and excellent research facilities greatly facilitated her contributions to this paper.

\section*{Appendix}
\subsection*{Appendix I}

We consider the spherically symmetric metric ansatz of the form
\begin{equation}
    \label{mm}
    ds^2=-\Phi(r) dt^2+ \frac{dr^2}{\Phi(r)}+r^2 d\Omega^2,
\end{equation}
where $d\Omega^2=d\theta^2+\sin^2(\theta) d\phi^2$ represents the line-element of unit 2-sphere.  One can find the non-zero components of the Ricci tensor,
Ricci scalar, and Einstein tensor as follows
\begin{eqnarray}
    &&R_0^0=R_1^1=-\frac{r\Phi_{,rr}+2\Phi_{,r}}{2r},\nonumber\\
    &&R_2^2=R_3^3=-\frac{r\Phi_{,r}+\Phi-1}{r^2},\nonumber\\
    &&R=-\frac{r^2\Phi_{,rr}+4r\Phi_{,r}+2\Phi-2}{r^2},\nonumber\\
    &&G_0^0=G_1^1=\frac{r\Phi_{,r}+\Phi-1}{r^2},\nonumber\\
    &&G_2^2=G_3^3=\frac{r\Phi_{,rr}+2\Phi_{,r}}{2r}.
  \end{eqnarray}
Consequently, the non-zero components of the Schouten tensors can be obtained
as\begin{eqnarray}{\label{Schouten}}
    &&\mathcal{S}_0^0=\mathcal{S}_1^1=-\frac{2r^2\Phi_{,rr}+5r\Phi_{,r}+\Phi-1}{3r^2},\\
    &&\mathcal{S}_2^2=\mathcal{S}_3^3=-\frac{r\Phi_{,r}+2\Phi-2}{3r^2}.
\end{eqnarray}
As the source of field equations,  one can consider the energy-momentum tensor $T^{\mu}_{\nu}$ with the following
non-zero components introduced by Kiselev \cite{Kiselev}
\begin{eqnarray}
    &&T^0_0=T^1_1=- \rho,\\
    &&T^2_2=T^3_3=\frac{(1+3\omega)}{2}\rho,
\end{eqnarray}
where $\omega$ is a constant  equation of state parameter, and $\rho=\rho(r)$
is the energy density function. Hence, the nonzero components of the generalized energy-momentum tensor ${\mathcal{T}^\mu}_\nu$ read as
\begin{eqnarray}
    &&\mathcal{T}_0^0=\mathcal{T}_1^1=-\frac{2+3\omega}{3}\rho,\\
    &&\mathcal{T}_2^2=\mathcal{T}_3^3=\frac{5+3\omega}{6}\rho.
\end{eqnarray}
The non-vanishing components of Cotton tensor $C_{\mu}^{\nu}$ can be obtained as
\begin{eqnarray}
    C^0_0&=&C^1_1=\mathcal{S}^0_0-\mathcal{T}^0_0=- \frac{2r^2\Phi_{,rr}+5r\Phi_{,r}+\Phi-1}{3r^2}+\frac{2+3\omega}{3}\rho,\\
    C^2_2&=&C^3_3=\mathcal{S}^2_2-\mathcal{T}^2_2=-\frac{r\Phi_{,r}+2\Phi-2}{3r^2}-\frac{5+3\omega}{6}\rho.
\end{eqnarray}
Hence,  substituting these in the Codazzi condition \eqref{codazzi2}, one finds the field equations as
\begin{eqnarray}
-2+2 \Phi+r\left[-2 \Phi_{,r}+r(-r(2+3\omega)\rho_{,r}+\Phi_{,rr}+r\Phi_{,rrr})\right]&=&0,\label{field eq 1}\\
2(-1+\Phi)+r\left[-2\Phi_{,r}+r(-9(1+\omega)\rho-r(5+3\omega)\rho_{,r}+\Phi_{,rr}
+\{\Phi_{,rrr})\right]&=&0\label{field eq 2}.
\end{eqnarray}
 Then, by using Eq. (\ref{field eq 1}), one can obtain $\rho_{,r}$ as
\begin{eqnarray}\label{rho,r}
    \rho_{,r}=\frac{1}{2+3\omega}\left(\frac{2\Phi-2}{r^3}-\frac{2\Phi_{,r}}{r^2}+\frac{\Phi_{,rr}}{r}+\Phi_{,rrr}\right),
\end{eqnarray}
where $\omega \neq -\frac{2}{3}$. For the ease of calculation, let us define $\gamma(r)=\frac{2\Phi-2}{r^3}-\frac{2\Phi_{,r}}{r^2}+\frac{\Phi_{,rr}}{r}+\Phi_{,rrr}$.
Hence, taking derivative of both sides of Eq.(\ref{rho,r}), we have
\begin{eqnarray}\label{(rho,rr)}
    \rho_{,rr}= \frac{1}{2+3\omega} \gamma_{,r},
\end{eqnarray}
where the function $\gamma_{,r}$ is
\begin{eqnarray*}
    \gamma_{,r}=-6 \frac{\Phi-1}{r^4}+6\frac{\Phi_{,r}}{r^3}-3 \frac{\Phi_{,rr}}{r^2}+ \frac{\Phi_{,rrr}}{r}+\Phi_{,rrrr}.
\end{eqnarray*}
Meanwhile, by Eq. (\ref{field eq 2}), one can obtain $\rho$ as
\begin{eqnarray}{\label{rho in 2}}
    \rho= \frac{r}{9(1+\omega)}(\gamma-(5+3\omega)\rho_{,r}),
\end{eqnarray}
where $\omega \neq -1$. Taking the derivative of both sides of Eq. (\ref{rho in 2}) reads
\begin{eqnarray}{\label{rho prime in 2}}
    \rho_{,r}&=& \frac{1}{9(1+\omega)}(\gamma-(5+3\omega)\rho_{,r})+ \frac{r}{9(1+\omega)}(\gamma_{,r}-(5+3\omega)\rho_{,rr}).
\end{eqnarray}
Now substituting Eq.(\ref{rho,r}) and Eq.(\ref{(rho,rr)}) into Eq.(\ref{rho prime in 2}) reads
\begin{eqnarray}{\label{solve}}
    \frac{1}{3(2+3\omega)(1+\omega)}\left[r\gamma_{,r}+\gamma(4+3\omega)\right]=0.
\end{eqnarray}
Hence, using the definition of $\gamma$ function, we have the following fourth
order nonhomogeneous ODE
for $\Phi$ function
\begin{eqnarray}\label{Euler-CauchyODE}
    \Phi_{,rrrr}r^4+(5+3\omega)\Phi_{,rrr}r^3+(1+3\omega)\Phi_{,rr}r^2-2(1+3\omega)\Phi_{,r}r+2(1+3\omega)\Phi=(2+6\omega).
\end{eqnarray}
Note that Eq. (\ref{Euler-CauchyODE}) is an Euler-Cauchy type nonhomogeneous ODE; therefore, one can find its complementary solution (to its associated homogeneous Euler ODE)
by taking $\Phi(r)=r^n$ where $n \in \mathbb{R}$. This gives the following polynomial
in $n$ \begin{eqnarray}
    (n-1)(n+1)(n-2)\left(n+(1+3\omega)\right)=0.
\end{eqnarray}
Consequently, the complementary solution $\Phi_c$ of Eq. (\ref{Euler-CauchyODE}) is
\begin{eqnarray}
    \Phi_c(r)=c_0r+\frac{c_1}{r}+c_2r^2+c_3r^{-(1+3\omega)},
\end{eqnarray}
where $c_0, c_1, c_2$ and $c_3$ are constants. Using the method of undetermined coefficients for the particular solution $\Phi_p$ to nonhomogeneous  Eq. (\ref{Euler-CauchyODE}), one obtains  $\Phi_p=1$. Hence we have
\begin{eqnarray}{\label{Kiselev general_a}}
    \Phi(r)=1+\frac{c_0}{r}+c_1r+c_2r^2+c_3r^{-(1+3\omega)},
\end{eqnarray}
 and the same energy-momentum tensor with the energy density  $\rho(r)$ as 
\begin{equation}
\rho(r)= 3\omega c_3 r^{-3 (1+\omega)},
\end{equation}
where $c_0, c_1, c_2$ and $c_3$ are arbitrary constants.
\subsection*{Appendix II}
For the metric \eqref{mm}, the Cotton, Ricci, Einstein, Schouten tensors and the Ricci scalar curvature function are the same as those in Appendix I. Considering the Dymnikova's energy-momentum tensor $T^\mu_\nu=\operatorname{diag}(-\rho, -\rho, p, p)$, where $\rho=\rho(r)$ and $p=p(r)$ are the energy density and pressure, respectively, one can find the corresponding generalized energy-momentum tensors in CT as
\begin{eqnarray}
    \mathcal{T}_0^0&=&\mathcal{T}_1^1=- \frac{2p + \rho}{3},\nonumber\\
    \mathcal{T}_2^2&=&\mathcal{T}_3^3= \frac{p+2 \rho}{3}.
\end{eqnarray}
Hence, the non-vanishing components of Cotton tensor $C^\mu_\nu$ can be found as 
\begin{eqnarray}
\mathcal{C}_0^0&=&\mathcal{C}_1^1=\mathcal{S}_0^0-\mathcal{T}_0^0
    = \frac{-1+\Phi-r(\Phi_{,r}+r\Phi_{,rr})}{3r^2}+ \frac{2p+\rho}{3},\nonumber\\
    \mathcal{C}_2^2&=&\mathcal{C}_3^3=\mathcal{S}_2^2-\mathcal{T}_2^2
    =\frac{4-4\Phi-2r\Phi_{,r}+r^2\Phi_{,rr}}{6r^2}-\frac{p+2\rho}{3}.
\end{eqnarray}
Hence, substituting these in the Codazzi condition \eqref{codazzi2}, one finds the field equations as
   \begin{eqnarray} &&-2+2\Phi+r\left(-2\Phi_{,r}+r(\Phi_{,rr}+r(-2p_{,r}-\rho_{,r}+\Phi_{,rrr}))\right)=0 \label{dym eq 1}\\
   &&-2+2\Phi-2 r\Phi_{,r}+r^2\Phi_{,rr}-2r^2(3p+3\rho+r(p_{,r}+2\rho_{,r}))+r^3\Phi_{,rrr}=0, \label{dym eq 2}
   \end{eqnarray}
   where $r\neq 0$. Then, by using Eq.(\ref{dym eq 1}), one can obtain $p_{,r}$ as
   \begin{eqnarray}{\label{p by 1}}
       p_{,r}=\frac{1}{2}(\gamma- \rho_{r}),
   \end{eqnarray}
   where $\gamma=\gamma(r)=\frac{2\Phi-2}{r^3}-\frac{2\Phi_{,r}}{r^2}+\frac{\Phi_{,rr}}{r}+\Phi_{,rrr}$. Meanwhile, by Eq.(\ref{dym eq 2}), one can obtain $p_{,r}$ as 
   \begin{eqnarray}\label{p by 2}
       p_{,r}=\frac{\gamma}{2}-\frac{3p+3\rho}{r}-2\rho_{,r}.
   \end{eqnarray}
Dymnikova's article gives the equation
\begin{eqnarray}{\label{-rho}}
    \rho(r)=\rho_{0}\operatorname{exp}(-\frac{r^3}{r_*^3}),
\end{eqnarray}
where $r_0=\sqrt{\frac{3}{\rho_0}}$ due to the de Sitter relation and $r_g$ is the Schwarzschild radius. Taking the derivative of both sides of Eq.(\ref{-rho}) with respect to $r$ reads
\begin{eqnarray}{\label{-rho_r}}
    \rho_{,r}=\rho_{0}\operatorname{exp}(-\frac{r^3}{r_*^3})(\frac{-3r^2}{r_*^3}).
\end{eqnarray}
Then, by Eqs. (\ref{p by 1})-(\ref{p by 2}), one can obtain $p_r$ as
\begin{eqnarray}
    p_{,r}&=&\frac{1}{2}\left(\gamma+( \frac{3r^2}{r_*^3}\rho_0)\operatorname{exp}(-\frac{r^3}{r_0^2r_g})\right)\label{abv 1}\\
    p_{,r}&=&\frac{\gamma}{2}-\frac{3p}{r}+\rho_0\operatorname{exp}(-\frac{r^3}{r_*^3})(\frac{6r^2}{r_*^3}-\frac{3}{r})\label{abv 2},
\end{eqnarray}
respectively.

\noindent Equating Eq.(\ref{abv 1}) and Eq.(\ref{abv 2}) gives
\begin{eqnarray}\label{p}
    p=\rho_0 \operatorname{exp}(-\frac{r^3}{r_*^3}) (\frac{3r^3-2r_*^3}{2r_*^3}). 
\end{eqnarray}
Taking the derivative of both sides of Eq.(\ref{p}) with respect to $r$ reads
\begin{eqnarray}{\label{p_r}}
    p_{,r}=\rho_0 \operatorname{exp}(-\frac{r^3}{r_*^3})(\frac{-9r^5+15r_*^3r^2}{2 r_*^6}).
\end{eqnarray}
Substituting Eqs.(\ref{-rho})-(\ref{-rho_r})-(\ref{p})-(\ref{p_r}) into Eqs.(\ref{dym eq 1})-(\ref{dym eq 2}) results in the following relation
\begin{eqnarray}{\label{dym res}}
     -2+2\Phi-2r\Phi_{,r}+r^2\Phi_{,rr}+r^3\Phi_{,rrr}+\frac{3 \rho_0\operatorname{exp}(\frac{-r^3}{r_*^3})r^5(3r^3-4r_*^3)}{r_*^6}=0.
\end{eqnarray}
Notice that Eq.(\ref{dym res}) is again an Euler-Cauchy type ODE. Therefore, we can find the solution using a similar method as in Appendix I. Then, the general solution of Eq.(\ref{dym res}) reads

\begin{eqnarray}
\Phi(r) = 1 + \frac{\rho_0 \exp\left(\frac{-r^3}{r_*^3}\right) r_*^3}{3r}+ \frac{c_0}{r} + c_1 r + c_2 r^2,
\end{eqnarray}
where \( c_0 \), \( c_1 \), and \( c_2 \) are constants. Defining $r_*^3= -r_0^2 c_0$ by the dimension analysis and using the de Sitter relation $r_0 = \sqrt{\frac{3}{\rho_0}}$, one can obtain the metric function as
\begin{eqnarray}\label{metdym}
\Phi(r) &=& 1 +  \frac{c_0}{r}\left( 1-\exp\left(-\frac{r^3}{r_*^3}\right) \right) + c_1 r  + c_2 r^2,\nonumber\\
&=&1 -  \frac{R_g(r)}{r} + c_1 r  + c_2 r^2
\end{eqnarray}
 and the same energy-momentum tensor with the energy density  $\rho(r)$ as 
\begin{eqnarray}
    \rho(r)=\rho_0\exp\left(-\frac{r^3}{r_*^3 }\right),
\end{eqnarray}
where $c_0, c_1, c_2, c_3, \rho_0$ are arbitrary constants and we defined $R_g(r)=-c_0\left( 1-\exp\left(-\frac{r^3}{r_*^3}\right) \right)$.\\

\newcommand{\bibTitle}[1]{``#1''}
\begingroup
\let\itshape\upshape
\bibliographystyle{plain}

\begin{thebibliography}{1}
\bibitem{Gurses}M. G\" urses, Y. Heydarzade, \c{C}. \c{S}ent{\" u}rk, \bibTitle{Wave Metrics in the Cotton and Conformal Killing Gravity Theories}, \href{ https://doi.org/10.1103/PhysRevD.110.084082}{Phys. Rev. D 110,  084082 (2024).}

\bibitem{Altas} E. Altas, B. Tekin, \bibTitle{Vanishing of Conserved Charges in Cotton Gravity}, \href{https://doi.org/10.48550/arXiv.2411.02132}{Phys.Rev.D 111, L021503, (2025).} 

\bibitem{Cotton} E. Cotton, \bibTitle{Sur les varietes a trois dimensions}, \href{https://afst.centre-mersenne.org/item/AFST_1899_2_1_4_385_0/}{Ann. Fac.
Sci. Toulouse 2. 1, 385 (1899).}

\bibitem{Weyl} H. Weyl, \bibTitle{Reine Infinitesimalgeometrie},
\href{https://doi.org/10.1007/BF01199420}{Math Z 2, 384-411 (1918).}

\bibitem{Eisenhart} L. P. Eisenhart, \bibTitle{Riemannian geometry}, Princeton University Press (1997),  ISBN 0-691-02353-0

\bibitem{deser}S. Deser, R. Jackiw, S.-Y. Pi, \bibTitle{Cotton blend gravity pp waves}, \href{https://www.actaphys.uj.edu.pl/R/36/1/27}{Acta Phys. Pol. B 36, 27 (2005).}

\bibitem{manheim1}
P. D. Mannheim and D. Kazanas, \bibTitle{Exact Vacuum Solution to Conformal Weyl Gravity and Galactic Rotation Curves}, \href{https://doi.org/10.1086/167623}{Astrophys. J. 342, 635 (1989).}

\bibitem{manheim2}P. D. Mannheim and D. Kazanas, \bibTitle{General Structure of the Gravitational Equations of Motion in Conformal Weyl Gravity}, \href{https://doi.org/10.1086/191573}{Astrophys. J. Suppl. Ser. 76, 431 (1991).}

\bibitem{Harada} H. Junpei, \bibTitle{Emergence of the Cotton tensor for describing gravity},  \href{https://doi.org/10.1103/PhysRevD.103.L121502}{Phys. Rev. D 103, L121502, (2021).}

\bibitem{Harada2} H. Junpei, \bibTitle{Cotton gravity and 84 galaxy rotation curves}, \href{https://doi.org/10.1103/PhysRevD.106.064044}{Phys. Rev. D 106, 064044, (2022).}

\bibitem{gen1}M. Gogberashvili and A. Girgvliani, \bibTitle{General spherically symmetric solution of Cotton gravity}, \href{https://doi.org/10.1088/1361-6382/ad1781}{Class. Quant. Gravity 41, 025010 (2024).}

\bibitem{gen2}Junior, E.L.B., Junior, J.T.S.S., Lobo, F.S.N. \it{et al.}, \bibTitle{Black bounces in Cotton gravity}, \href{https://doi.org/10.1140/epjc/s10052-024-13568-x}{Eur. Phys. J. C 84, 1190 (2024).}

\bibitem{gen3}R. A. Sussman, S. Najera, \bibTitle{Exact solutions of Cotton Gravity in its Codazzi formulation}, \href{https://doi.org/10.48550/arXiv.2312.02115}{arXiv:2312.02115 [gr-qc].}

\bibitem{crit1}P. Bargueno, \bibTitle{Comment on Emergence of the Cotton tensor for describing gravity}, \href{https://doi.org/10.1103/PhysRevD.104.088501}{Phys. Rev. D 104, 088501 (2021).}

\bibitem{crit2} J. Harada, \bibTitle{Reply to Comment on Emergence of the Cotton tensor for describing gravity,} \href{https://doi.org/10.1103/PhysRevD.104.088502}{Phys. Rev. D 104, 088502 (2021).}

\bibitem{crit3} G. Clement and K. Nouicer, \bibTitle{Cotton gravity is not predictive}, \href{https://doi.org/10.1016/j.physletb.2024.138947 }{Phys. Lett. B 856, 138947 (2024).}

\bibitem{crit4} R. A. Sussman, C. A. Mantica, L.G. Molinari, and S. Najera, \bibTitle{Response to a critique of "Cotton Gravity"}, \href{https://doi.org/10.48550/arXiv.2401.10479}{arXiv:2401.10479 [gr-qc].}

\bibitem{crit5} G. Clement and K. Nouicer, \bibTitle{Farewell to Cotton gravity}, \href{https://doi.org/10.48550/arXiv.2401.16008}{arXiv:2401.16008 [gr-qc].}

\bibitem{crit6} R. A. Sussman, C. A. Mantica, L. G. Molinari, and S. Najera, \bibTitle{Second Response to the critique of "Cotton Gravity"}, \href{https://doi.org/10.48550/arXiv.2402.01992}{arXiv:2402.01992 [gr-qc].}

\bibitem{Mantica} C. A. Mantica and L. G. Molinari, \bibTitle{Codazzi tensors and their space-times and cotton gravity},
\href{https://doi.org/10.1007/s10714-023-03106-7}{Gen. Relativ. Gravit, 55,
62, (2023).}

\bibitem{Kiselev} V. V. Kiselev, \bibTitle{Quintessence and black holes}, \href{https://doi.org/10.1088/0264-9381/20/6/310}{Class. Quant. Gravity 20, 1187 (2003).}

\bibitem{Dymnikova} I. Dymnikova, \bibTitle{Vacuum nonsingular black hole}, \href{https://doi.org/10.1007/BF00760226}{Gen. Relativ. Gravit 24, 235242, (1992).}

\bibitem{yh1} Y. Heydarzade, F. Darabi, \bibTitle{Surrounded Vaidya solution by cosmological fields}, \href{https://doi.org/10.1140/epjc/s10052-018-6041-4}{Eur. Phys. J. C 78, 582 (2018).}

\bibitem{yh2} Y. Heydarzade, F. Darabi, \bibTitle{Surrounded Vaidya black holes: apparent horizon properties}, \href{https://doi.org/10.1140/epjc/s10052-018-5842-9}{Eur. Phys. J. C 78, 342 (2018).}

\bibitem{yh3}H. Hadi, F. Darabi, K. Atazadeh, and Y. Heydarzade, \bibTitle{D-bound and the Bekenstein bound for the surrounded Vaidya black hole}, \href{https://doi.org/10.1140/epjc/s10052-020-08699-w}{Eur. Phys. J. C 80, 1126 (2020).}

\bibitem{yh4}H. Hadi, Y. Heydarzade, F. Darabi, and K. Atazadeh, \bibTitle{D-bound and Bekenstein bound for McVittie solution surrounded by dark energy cosmological fields}, \href{https://doi.org/10.1140/epjp/s13360-020-00601-7}{Eur. Phys. J. Plus 135, 584 (2020).}

\bibitem{bh1} R. Geroch and J.B. Hartle, \bibTitle{Distorted black holes}, \href{https://doi.org/10.1063/1.525384}{J. Math. Phys. 23, 680 (1982).}

\bibitem{bh2} S. Fairhurst, B. Krishnan, \bibTitle{Distorted Black Holes with Charge}, \href{https://doi.org/10.1142/S0218271801001086}{Int. J. Mod. Phys. D10, 691 (2001).}

\bibitem{bh3}S. R. Brandt, E. Seidel, \bibTitle{The Evolution of Distorted Rotating Black Holes III: Initial Data}, \href{https://doi.org/10.1103/PhysRevD.54.1403}{Phys. Rev. D 54, 1403 (1996).}

\bibitem{bh4} M. Ansorg, D. Petroff, \bibTitle{Black holes surrounded by uniformly rotating rings}, \href{https://doi.org/10.1103/PhysRevD.72.024019}{Phys. Rev. D 72, 024019 (2005).}

\bibitem{bh5}S. W. Hawking, \bibTitle{Black holes in general relativity}, \href{https://doi.org/10.1007/BF01877517}{Commun. Math. Phys. 25, 152 (1972).}

\bibitem{bh6} J. D. Brown and V. Husian, \bibTitle{Black holes with short hair}, \href{https://doi.org/10.1142/S0218271897000340}{Int. J. Mod. Phys. D 6, 563 (1997).}

\bibitem{bh7} S.Droz, M. Heusler, N. Straumann, \bibTitle{New black hole solutions with hair}, \href{https://doi.org/10.1016/0370-2693(91)91592-J}{Phys. Lett. B, 268, 371 (1991).} 

\bibitem{bh8} J. Barranco, A. Bernal, J.C. Degollado, A. Diez-Tejedor, M. Megevand, M. Alcubierre, D.Nunez, O. Sarbach, \bibTitle{Schwarzschild Black Holes can Wear Scalar Wigs}, \href{https://doi.org/10.1103/PhysRevLett.109.081102}{Phys. Rev. Lett. 109, 081102 (2012).}

\bibitem{reg1}K. A. Bronnikov, \bibTitle{Regular magnetic black holes and monopoles from nonlinear electrodynamics}, \href{https://doi.org/10.1103/PhysRevD.63.044005}{Phys.
Rev. D 63, 044005 (2001).}

\bibitem{reg2}G. G. L. Nashed, \bibTitle{Vacuum nonsingular black hole solutions in tetrad theory of gravitation},
\href{https://doi.org/10.1023/A%3A1016509920499}{Gen. Rel. Grav. 34, 1047-1058 (2002).}

\bibitem{reg3} S. Shankaranarayanan, N. Dadhich, \bibTitle{Nonsingular black holes on the brane}, \href{https://doi.org/10.1142/S0218271804005109}{Int.
J. Mod. Phys. D 13, 1095-1104 (2004).}

\bibitem{reg4}I. Dymnikova, \bibTitle{Regular electrically charged structures in nonlinear electrodynamics coupled to general relativity}, \href{https://doi.org/10.1088/0264-9381/21/18/009}{Class.
Quant. Grav 21, 4417 (2004).}

\bibitem{reg5}S. A. Hayward, \bibTitle{Formation and evaporation of regular black holes}, \href{https://doi.org/10.1103/PhysRevLett.96.031103}{Phys.
Rev. Lett. 96, 031103 (2006).}

\bibitem{reg6}K.A. Bronnikov, \bibTitle{Regular phantom black holes}, \href{https://doi.org/10.1103/PhysRevLett.96.251101}{Phys.
Rev. Lett. 96, 251101 (2006).}

\bibitem{reg7}W. Berej, J. Matyjasek, D. Tryniecki, M. Woronowicz, \bibTitle{Regular black holes in quadratic gravity}, \href{https://doi.org/10.1007/s10714-006-0270-9}{Gen.
Rel. Grav. 38, 885 (2006).}

\bibitem{reg8}I. Dymnikova, E. Galakatinov, \bibTitle{Regular rotating de Sitter-Kerr black holes and solitons}, \href{https://doi.org/10.1088/0264-9381/33/14/145010}{Class.
Quant. Grav. 33,  145010 (2016).}

\bibitem{reg9}V. P. Frolov, \bibTitle{Notes on nonsingular models of black holes}, \href{https://doi.org/10.1103/PhysRevD.94.104056}{Phys. Rev. D 94, 104056 (2016).} 
\bibitem{1a} J. Matyjasek, and M. Telecka, \bibTitle{Kiselev and Schwarzschild-de Sitter black holes in higher derivative theories of gravitation}, \href{https://doi.org/10.1103/PhysRevD.107.064058}{Phys. Rev. D 107, 064058 (2023)}.
\bibitem{2a}
Ma. Dariescu, C. Dariescu, V. Lungu, and C. Stelea, \bibTitle{Kiselev solution in power-Maxwell electrodynamics}, \href{https://doi.org/10.1103/PhysRevD.106.064017}{Phys. Rev. D 106, 064017 (2022)}.
\bibitem{3a}M. E. Rodrigues, E. L. B. Junior, G. T. Marques, and V. T. Zanchin, \bibTitle{Regular black holes in $f(R)$ gravity coupled to nonlinear electrodynamics}, \href{https://doi.org/10.1103/PhysRevD.94.024062} {Phys. Rev. D 94, 024062 (2016)}.
\bibitem{4a} S. Nojiri, S. D. Odintsov, \bibTitle{Regular multihorizon black holes in modified gravity with nonlinear electrodynamics}, \href{https://doi.org/10.1103/PhysRevD.96.104008}{Phys. Rev. D 96, 104008 (2017)}.
\bibitem{5a} S. G. Ghosh, D. V. Singh, and S. D. Maharaj, \bibTitle{Regular black holes in Einstein-Gauss-Bonnet gravity}, \href{https://doi.org/10.1103/PhysRevD.97.104050}{Phys. Rev. D 97, 104050 (2018)}.
\bibitem{6a}L. Heisenberg, R. Kase, M. Minamitsuji, and S. Tsujikawa,  \bibTitle{Hairy black-hole solutions in generalized Proca theories}, \href{https://doi.org/10.1103/PhysRevD.96.084049}{Phys. Rev. D 96, 084049 (2017)}.
\bibitem{7a} M. E. Rodrigues and M. V. de S. Silva, \bibTitle{Regular multihorizon black holes in $f(G)$ gravity with nonlinear electrodynamics}, \href{https://doi.org/10.1103/PhysRevD.99.124010}{Phys. Rev. D 99, 124010 (2019)}.

\bibitem{m-mantica} C. A. Mantica, and L. G. Molinari,
\bibTitle{Friedmann equations in the Codazzi parametrization of Cotton and extended theories of gravity and the Dark Sector}, \href{https://doi.org/10.1103/PhysRevD.109.044059}{Phys. Rev. D 109, 044059 (2024)}.
    
\bibitem{Petrov}A. Z. Petrov, \bibTitle{Einstein spaces}, Pergamon Press, Oxford, (1969), ISBN 10: 0080123155.

\bibitem{Mantica2023} C. A. Mantica, and L. G. Molinari, ``The covariant approach to the static spacetimes in Einstein and extended gravity theories'', \href{https://doi.org/10.1007/s10714-023-03149-w}{Gen. Relativ. Gravit. \textbf{55}, 100 (2023)}.

\bibitem{r41}  P. Berglund, J. Bhattacharyya, and D. Mattingly, \bibTitle{Mechanics of universal horizons}, \href{https://doi.org/10.1103/PhysRevD.85.124019}{Phys. Rev. D 85, 124019 (2012).}

\bibitem{r42} M. G\" urses, E. Sermutlu, \bibTitle{Static spherically symmetric solutions to Einstein-Maxwell dilaton field equations in D dimensions}, \href{https://doi.org/10.1088/0264-9381/12/11/011}{Class. Quant. Grav. 12, 2799 (1995).}

\bibitem{heydar} M. G\" urses, Y. Heydarzade, \c{C}. \c{S}ent{\" u}rk, \bibTitle{NAT black holes}, \href{https://doi.org/10.1140/epjc/s10052-019-7455-3}{Eur. Phys. J. C 79, 942 (2019).}


\bibitem{Komar} A. Komar, \bibTitle{Positive-Definite Energy Density and Global Consequences for General Relativity}, \href{https://doi.org/10.1103/PhysRev.129.1873}{Phys. Rev. 129, 1873 (1963)}.

\bibitem{ADM} R. Arnowitt, S. Deser, C. W. Misner, \bibTitle{The dynamics of general relativity}, In L. Witten, editor, Gravitation: An Introduction to Current Research, Wiley, (1962).

\bibitem{Bekenstein} J. D. Bekenstein, \bibTitle{Black Holes and Entropy}, \href{https://doi.org/10.1103/PhysRevD.7.2333}{Phys. Rev. D 7, 2333 (1973).}

\bibitem{Hawking} S. W. Hawking, \bibTitle{Particle creation by black holes}, \href{https://doi.org/10.1007/BF02345020}{Commun. Math. Phys. 43, 199 (1975).}


\end{thebibliography}

\end{document}